\documentclass[%
 reprint,
 amsmath,amssymb,
 aps,
 prb,
]{revtex4-1}

\usepackage{dcolumn}% Align table columns on decimal point
\usepackage{bm}% bold math
    \usepackage{amsmath,amssymb, mathrsfs}
\usepackage[colorlinks=true,bookmarks=false,citecolor=blue,urlcolor=blue]{hyperref} %pdflatex

\usepackage{graphicx}
\usepackage{float}
\usepackage{xcolor}
\usepackage{caption}
\begin{document}

\title{Topological Temporally Mode-Locked Laser}

\author{Christian R. Leefmans$^{1,\ast}$, Midya Parto$^{2}$, James Williams$^{2}$,\\Gordon H.Y. Li$^{1}$, Avik Dutt$^{3,4}$, Franco Nori$^{5,6,7}$ and Alireza Marandi$^{1,2,\dagger}$\\
\textit{$^1$Department of Applied Physics, California Institute of Technology, Pasadena, CA 91125, USA.\\
$^2$Department of Electrical Engineering, California Institute of Technology,Pasadena, CA 91125, USA.\\
$^3$Department of Electrical Engineering, Stanford University,Stanford, CA 94305, USA.\\
$^4$Department of Mechanical Engineering and IPST, University of Maryland, College Park, MD 20742, USA.\\
$^{5}$Theoretical Quantum Physics Laboratory, RIKEN Cluster for Pioneering Research,Wako, Saitama 351-0198, Japan.\\
$^{6}$RIKEN Center for Quantum Computing, Wako, Saitama 351-0198, Japan.\\$^{7}$Department of Physics, University of Michigan, Ann Arbor, Michigan 48109-1040, USA.}\\
$^\ast$\href{mailto:cleefman@caltech.edu}{cleefman@caltech.edu}, $^\dagger$\href{mailto:marandi@caltech.edu}{marandi@caltech.edu}
}

\date{\today}

\maketitle

\textbf{Mode-locked lasers play a key role in modern science and technology. Not only do they lay the foundation for ultrafast optics and play a central role in nonlinear optics, but also they have important applications in imaging~\cite{helmchen_deep_2005}, metrology~\cite{udem_optical_2002}, telecommunications~\cite{keller_recent_2003}, sensing~\cite{picque_frequency_2019}, and computing~\cite{marandi_network_2014}. While substantial efforts have focused on mode-locking the spectral modes of lasers~\cite{haus_mode-locking_2000}, relatively little attention has been paid to mode-locking their temporal modes. However, temporal mode-locking presents ample opportunities to develop new technologies~\cite{cole_soliton_2017} and to study the intersection of nonlinear, non-Hermitian, and topological phenomena~\cite{herink_real-time_2017}, which, in recent years, has been a priority for the field of topological physics~\cite{smirnova_nonlinear_2020,bergholtz_exceptional_2021,parto_non-hermitian_2021}. Here, we theoretically predict and experimentally realize \textit{topological temporal mode-locking} in a laser cavity with time-delayed intracavity couplings. These couplings introduce non-Hermitian point-gap topology~\cite{gong_topological_2018} into the temporal modes of our laser and conspire with an effective nonlocal nonlinearity to generate mode-locked temporal structures in our laser cavity, whose form can be tailored by suitably engineering the underlying couplings. We harness this approach to realize a nonlinearity-driven non-Hermitian skin effect~\cite{gong_topological_2018}, and we show that the topological temporal modes of our laser are robust against disorder-induced localization~\cite{hatano_non-hermitian_1998}. The flexibility and programmability of our topological temporal mode-locking scheme reveals new opportunities to study nonlinear~\cite{smirnova_nonlinear_2020} and non-Hermitian~\cite{bergholtz_exceptional_2021} topological phenomena in mode-locked photonic resonators and could enable new applications of mode-locked lasers to sensing and optical computing.}

Mode-locking occurs in a photonic resonator when nonlinear interactions lock the amplitudes and phases of the resonator's modes relative to one another~\cite{wright_mechanisms_2020}. This process occurs in a variety of photonic systems, including lasers \cite{haus_mode-locking_2000} and Kerr cavities~\cite{leo_temporal_2010}, and it plays a crucial role in existing and emerging technologies. Most research efforts in this area have focused on \textit{spectral} mode-locking, in which a nonlinear mechanism, such as active modulation~\cite{haus_mode-locking_2000}, saturable absorption~\cite{keller_semiconductor_1996}, or four-wave mixing~\cite{kippenberg_dissipative_2018}, synchronizes the longitudinal, or spectral, modes of a resonator. However, recent work suggests that \textit{spatial} and \textit{temporal} mode-locking can produce novel phenomena that are interesting from both fundamental and practical perspectives~\cite{tegin_spatiotemporal_2019,wright_spatiotemporal_2017,cole_soliton_2017,wang_optical_2019}. For instance, recent results suggest that temporal mode-locking can provide new tools for telecommunications~\cite{lu_synthesized_2021}, photonic computing~\cite{marandi_network_2014}, all-optical data storage~\cite{cole_soliton_2017}, and the study of topological physics~\cite{herink_real-time_2017}. Yet, despite this promising potential, a scheme to controllably and programmably mode-lock the temporal modes of a laser cavity has not been realized.

Meanwhile, a principal objective of topological physics is to understand the interplay between nonlinearity, non-Hermiticity, and topology~\cite{smirnova_nonlinear_2020,bergholtz_exceptional_2021,parto_non-hermitian_2021}. While photonic experiments on nonlinear and non-Hermitian topology have largely utilized passive optical nonlinearities~\cite{mukherjee_observation_2020,jurgensen_quantized_2021} and linear gain and loss~\cite{weimann_topologically_2017,zhao_non-hermitian_2019}, recent theoretical studies on mode-locked topological resonators point to their unprecedented potential to realize novel nonlinear and non-Hermitian topological effects~\cite{yang_mode-locked_2020,mittal_topological_2021,tikan_protected_2022}.  However, despite a growing number of such theoretical proposals~\cite{longhi_non-hermitian_2019,zykin_topological_2021,tusnin_dissipative_2021}, experimental realizations of topological mode-locked resonators remain rare. In addition, existing proposals for mode-locked topological resonators largely rely on topological mode-locking in the spectral domain, which can limit the study of topological edge effects or the realization of topological lattice models with inhomogeneous couplings in mode-locked systems. The dearth of experimental topological mode-locked resonators and the restrictions of topological mode-locking in the frequency domain strongly motivate the search for an alternative mode-locking scheme that can straightforwardly and controllably introduce topology into mode-locked resonators.

\begin{figure*}
    \centering
    \includegraphics[width=\textwidth]{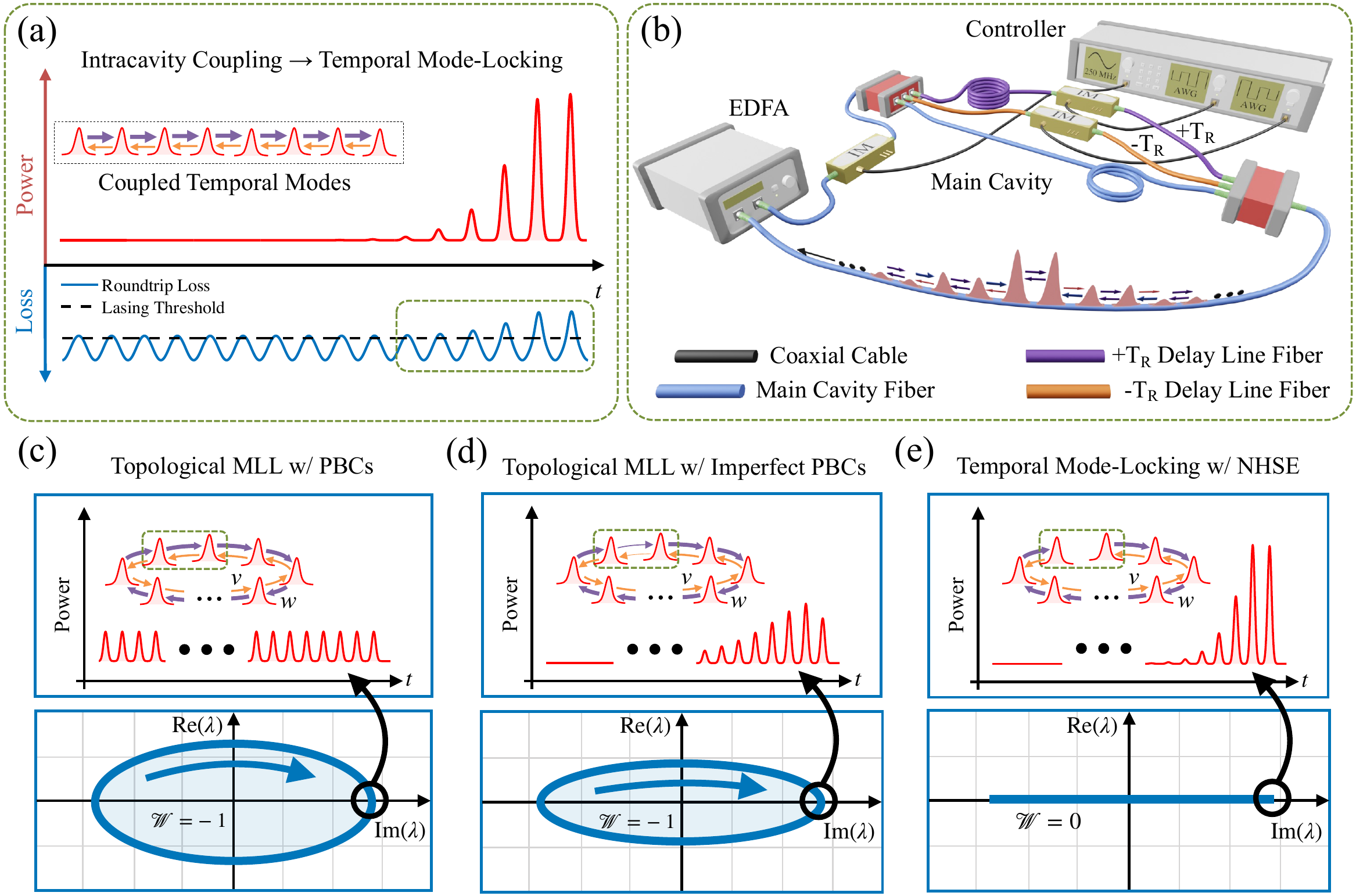}
\caption{{\textbf{Topological Temporal Mode-Locking}}~{\textbf{(a)}}~General concept of the topological temporally mode-locked laser (TTMLL). Sinusoidal modulation of the intracavity loss produces $N$ evenly spaced temporal modes in the laser cavity. Dissipative coupling between the modes alters the roundtrip loss of the modes and, together with the slow gain saturation nonlinearity, gives rise to a mode-locked temporal structure. {\textbf{(b)}}~Schematic of the fiber-based TTMLL used in our experiments. The architecture consists of a main laser cavity with two optical delay lines, which introduce dissipative couplings between nearest-neighbor pulses in the cavity. EDFA: Erbium-doped fiber amplifier. IM: Intensity Modulator. {\textbf{(c-e)}}~Schematic illustration of topological temporal mode-locking with the Hatano-Nelson (HN) model, which is a one-dimensional chain with asymmetric couplings $w$ and $v$. $\mathcal{W}$ represents the topological winding number. Note that, due to the nature of dissipative couplings, $\operatorname{Im}(\lambda)$ appears on the abscissa while $\operatorname{Re}(\lambda)$ appears on the ordinate~\cite{leefmans_topological_2022}.}
    \label{fig:schematic}
\end{figure*}

\begin{figure*}
    \centering
    \includegraphics[width=\textwidth]{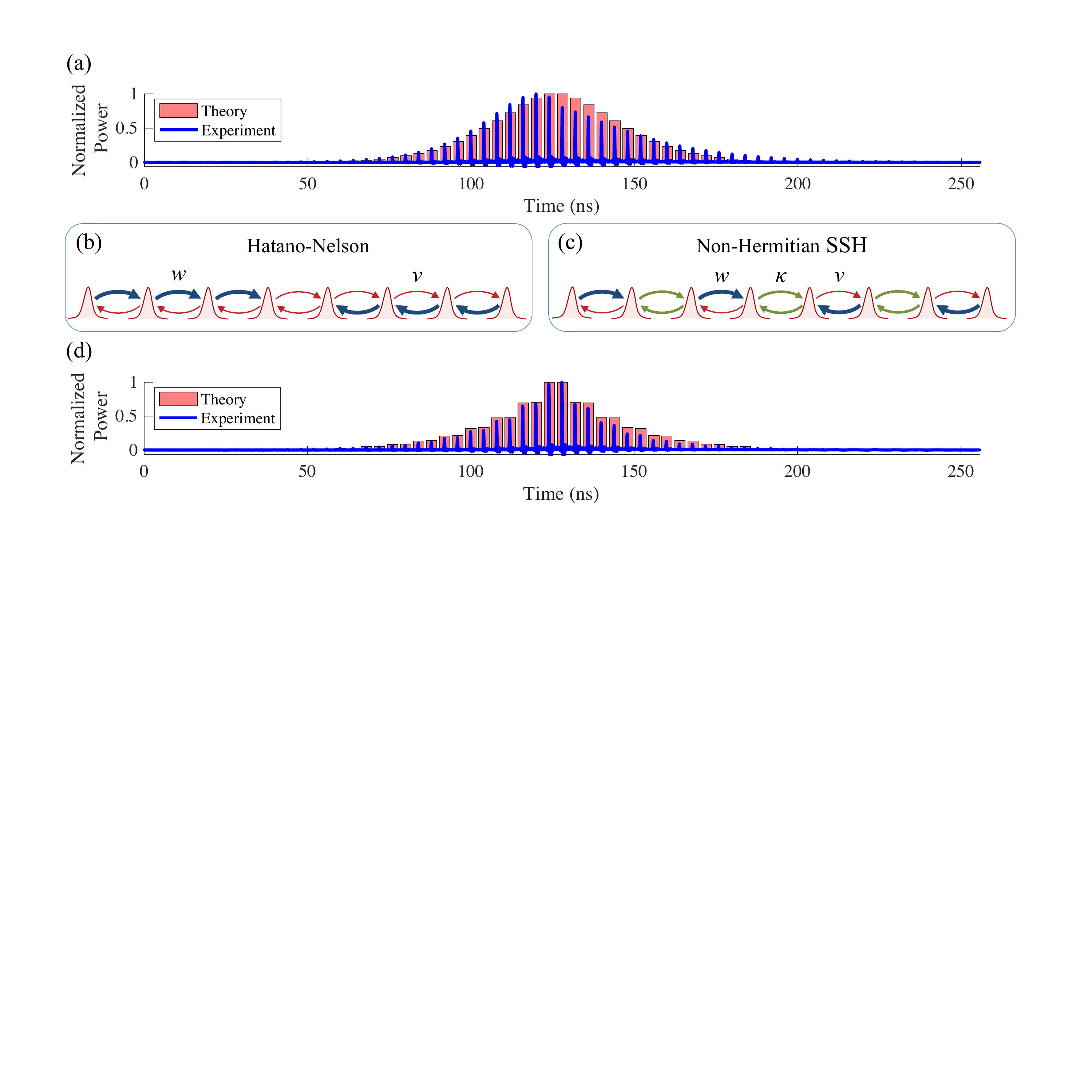}
    \caption{\textbf{Topological Mode-Locking at HN and NH-SSH Domain Walls}~\textbf{(a)}~Topologically mode-locked domain-wall skin mode corresponding to an HN domain wall. \textbf{(b)}~Schematic representation of the HN domain wall implemented for the measurement in \textbf{(a)}.  \textbf{(c)}~Schematic representation of the NH-SSH domain wall implemented for the measurement in \textbf{(d)}. \textbf{(d)}~Topologically mode-locked domain-wall skin mode corresponding to an NH-SSH domain wall. Note that for our experiments we choose $w/v=\sqrt{2}$ and $\kappa = \left(w+v\right)/2$.}
    \label{fig:domain_walls}
\end{figure*}

In this work, we introduce and experimentally realize a scheme to topologically mode-lock a laser's temporal modes, which we refer to \textit{topological temporal mode-locking}. We begin by introducing the laser architecture that underlies this scheme, the topological temporally mode-locked laser (TTMLL). We then utilize this laser architecture to experimentally demonstrate a nonlinearity-driven non-Hermitian skin effect (NHSE)~\cite{gong_topological_2018} and to demonstrate the topological robustness of our mode-locked laser's (MLL's) temporal modes against disorder-induced localization. Our work demonstrates a flexible and programmable temporal mode-locking scheme that is capable of implementing arbitrary couplings between the temporal modes of our MLL~\cite{leefmans_topological_2022}. This capability not only enables us to study diverse topological lattice models under different boundary conditions, but it may also enable novel applications of MLLs to sensing and optical computing. Moreover, the temporal mode-locking scheme discussed in this work may be straightforwardly extended to other photonic resonators, which exposes a bevy of opportunities to explore nonlinear and non-Hermitian topology in mode-locked photonic resonators.

The dynamics of the TTMLL can be described by a modified master equation for active mode-locking~\cite{haus_mode-locking_2000}. In the low power regime, in which the Kerr nonlinearity may be neglected, this master equation models $N$ evenly spaced temporal modes (laser pulses), which are strongly localized within their respective ``gain wells'' [see the correspondence between the pulse positions and loss minima in Fig.~\ref{fig:schematic}\textcolor{red}{(a)}]. Due to this strong localization, we can approximate the master equation by a tight-binding model of the form [see Supplementary Information Sec.~7]:

\begin{subequations}
\begin{equation}
    \frac{\partial{a_{n}}}{\partial{T}} = \left[K^{L}_{n}a_{n-1} + K^{R}_{n}a_{n+1}\right] + \left(g(T)-\Gamma\right)a_{n}
    \label{eq:tight_binding_eq}
\end{equation}
    
\begin{equation}
    \frac{\partial{g}}{\partial{T}}=-\gamma\left(g-g_{0}+g\epsilon\sum_{n}\left|a_{n}(T)\right|^{2}\right).
    \label{eq:tight_binding_gain}
\end{equation}
\label{eq:tight_binding_model}
\end{subequations}

\noindent Here $a_{n}$ is the amplitude of the $n{\rm th}$ pulse, $g(T)$ describes the nonlinear laser gain, $g_{0}$ is the small signal gain, $\Gamma$ is the effective linear loss, $\gamma=T_{\rm RT}/\tau$ is the ratio of the roundtrip period and the gain relaxation time, and $\epsilon$ is related to the saturation intensity. The terms in square brackets in Eq.~\ref{eq:tight_binding_eq} describe nearest-neighbor intracavity couplings between the pulses. We introduce non-Hermitian topology into our TTMLL by engineering $K_{n}^{L}$ and $K_{n}^{R}$ to implement lattice models that exhibit non-Hermitian point gap topology~\cite{gong_topological_2018}. While we only consider nearest-neighbor couplings in Eq.~\ref{eq:tight_binding_eq}, it is possible to introduce arbitrary couplings between the pulses~\cite{leefmans_topological_2022}.

The interplay between the topological intracavity couplings and the gain saturation nonlinearity produces topological temporal mode-locking in our laser. As is shown in Fig.~\ref{fig:schematic}\textcolor{red}{(a)}, the intracavity dissipative couplings modify the loss landscape experienced by the laser's temporal modes. Meanwhile, the laser gain, whose dynamics we assume to be slow ($\gamma\ll1$), acts as a source of nonlocal nonlinear interactions between the pulses, in the sense that the gain saturates due to the average power in the laser cavity [see Eq.~\ref{eq:tight_binding_gain}]. These nonlinear interactions enable the pulses that experience less linear loss due to the dissipative couplings to ``consume'' more gain than the other pulses, and this drives all of the pulses collectively towards the lowest-loss temporal structure. Indeed, our simulations and analysis in Supplementary Information Secs.~8 and 9 indicate that the system stabilizes in the temporal structure that corresponds to the lowest-loss right eigenstate defined by the intracavity dissipative couplings. Because the dynamics of the TTMLL fixes the relative phases and amplitudes of the pulses inside the cavity, the combination of the non-Hermitian topological couplings and slow gain saturation leads to topological temporal mode-locking.

We present an example of topological temporal mode-locking in Figs.~\ref{fig:schematic}\textcolor{red}{(c-e)}, where we illustrate topological temporal mode-locking with the Hatano-Nelson (HN) model~\cite{hatano_non-hermitian_1998,hatano_vortex_1997}, which consists of a one-dimensional chain with asymmetric couplings $w$ and $v$. When our intracavity couplings implement the HN model with periodic boundary conditions (PBCs), the HN lattice exhibits a nontrivial topological winding number $\mathcal{W}$ in the complex energy plane~\cite{kawabata_symmetry_2019}, and the mode-locked temporal structure in our TTMLL is evenly distributed among the laser pulses [Fig.~\ref{fig:schematic}\textcolor{red}{(c)}]. When we reduce the coupling at the boundary of the lattice [Fig.~\ref{fig:schematic}\textcolor{red}{(d)}], the topological winding number remains nontrivial [$\mathcal{W}\neq0$], but the steady-state temporal structure in our laser begins to localize near one boundary. This localization is accompanied by an increase in the laser's peak power because the energy provided by our gain medium becomes concentrated in a smaller number of pulses. Finally, in the presence of open boundary conditions (OBCs) [Fig.~\ref{fig:schematic}\textcolor{red}{(e)}], the eigenvalues of the HN model collapse onto a line, and the lattice is topologically trivial [$\mathcal{W}=0$]~\cite{gong_topological_2018}. In this limit, the pulses in our TTMLL experience a nonlinearity-driven NHSE, and the mode-locked temporal structure is strongly localized near the boundary. It is important to note that, although this NHSE occurs in the topologically trivial phase, it is a distinctly topological phenomenon, in the sense that it both guarantees and is guaranteed by the presence of a topological invariant in the bulk lattice~\cite{zhang_correspondence_2020}. 

While the discussion above describes the dynamics of the TTMLL at low powers, at higher powers the Kerr nonlinearity can alter the phases, intensities, and pulse widths of the mode-locked topological temporal structure in the TTMLL. In the present experiments, we operate in a regime in which the Kerr nonlinearity has little effect on the intensity distribution of the observed temporal structures, which allows us to compare the measured intensities to the predictions of Eq.~\ref{eq:tight_binding_model}. We consider the effect of the Kerr nonlinearity in simulations presented in Supplementary Information Sec.~8.

We experimentally realize a topological temporally mode-locked laser by constructing the fiber-based TTMLL shown in Fig.~\ref{fig:schematic}\textcolor{red}{(b)}. This laser consists of a main laser cavity and two delay lines [labeled $\pm T_{\rm R}$]. We perform active mode-locking by inserting an erbium-doped fiber amplifier (EDFA) and an intensity modulator (IM) into the main cavity. The EDFA provides a slow gain medium ($\gamma=T_{\rm RT}/\tau\approx10^{-5}$), whose gain saturates due to the average power in the main cavity. Meanwhile, sinusoidally modulating the intracavity IM at a frequency of $f_{\rm mod}=64/T_{\rm RT}$ gives rise to $N=64$ time-multiplexed actively mode-locked pulses with widths of $\sim\!100~{\rm ps}$ and a repetition period of $T_{\rm rep}\approx 4~{\rm ns}$ [see Supplementary Information Sec.~3]. The two delay lines implement the nearest-neighbor dissipative couplings described in Eq.~\ref{eq:tight_binding_eq}, while the IMs in the delay lines enable us to modulate the strengths of these couplings both within one roundtrip and between roundtrips. Note that the two delay lines enable us to control each direction of the nearest-neighbor couplings independently~\cite{leefmans_topological_2022}. This facilitates implementing non-Hermitian topological lattice models in our system.

For our first demonstration of topological mode-locking, we program the dissipative couplings of our TTMLL to realize the HN domain wall shown in Fig.~\ref{fig:domain_walls}\textcolor{red}{(b)}. In this instance, we expect the NHSE to localize all modes of the system at the domain wall between the inverted HN domains, which possess different bulk topological winding numbers~\cite{gong_topological_2018}. To observe this localization experimentally, we begin by tuning our laser to operate with a small signal gain of $\sim\!23.7~{\rm dB}$, which, in the presence of the HN domain wall, results in an average output power of $\sim\!40~{\rm \mu{W}}$. We then switch the dissipative couplings of our laser to implement the HN domain wall with a coupling ratio of $w/v=\sqrt{2}$. Shortly thereafter, we observe that the pulse pattern in the laser becomes temporally mode-locked in a skin mode at the domain wall of the synthetic lattice [see Fig.~\ref{fig:domain_walls}\textcolor{red}{(a)}]. The intensity distribution of this domain-wall skin mode is in excellent agreement with the theoretically predicted lowest-loss skin mode of the underlying HN lattice. 

To showcase the flexibility of our topological mode-locking scheme, we repeat our domain wall localization experiment with the non-Hermitian Su-Schrieffer-Heeger (NH-SSH)~\cite{su_solitons_1979,yin_geometrical_2018,weidemann_topological_2020} domain wall shown in Fig.~\ref{fig:domain_walls}\textcolor{red}{(c)}. Here, we program the dissipative couplings of our laser to implement the NH-SSH domain wall with $w/v=\sqrt{2}$ and $\kappa=\left(w+v\right)/2$, and we operate our laser with a small signal gain of $\sim\!23.1~\rm{dB}$, where the average output power in the presence of the NH-SSH domain wall is $\sim\!42~ {\rm \mu{W}}$. Once again, we find that the intensity of the measured domain-wall skin mode is in excellent agreement with the theoretical lowest-loss skin mode of the underlying lattice. Notably, we are clearly able to distinguish the the NH-SSH skin mode, with its pairwise amplitudes [see Fig.~\ref{fig:domain_walls}\textcolor{red}{(d)}], from the HN skin mode observed in Fig.~\ref{fig:domain_walls}\textcolor{red}{(a)}, which demonstrates that the observed skin mode indeed reflects the couplings of the underlying lattice.

While domain-wall localization has been observed previously in linear point-gap topological systems, where it is known as topological funneling~\cite{weidemann_topological_2020}, we emphasize that the dynamics responsible for domain wall localization in our topological MLL is fundamentally different. Topological funneling also utilizes the NHSE, but it is a linear transport phenomenon, in which excitations in a non-Hermitian lattice localize at a domain wall due to an effective asymmetry in the system's conservative couplings~\cite{weidemann_topological_2020}. In contrast, domain wall localization that occurs in our topological MLL is not a transport phenomenon and is instead driven by the combination of asymmetric dissipative couplings and a nonlocal gain saturation nonlinearity.  Moreover, this process occurs in the absence of externally introduced excitations due to the presence of net gain in our topological MLL, which enables the observed domain-wall skin modes to form from noise in the laser cavity. In this sense, the domain wall localization in our topological MLL is a nonlinearity-driven NHSE and is fundamentally distinct from topological funneling.

\begin{figure*}
    \centering
    \includegraphics[width=\textwidth]{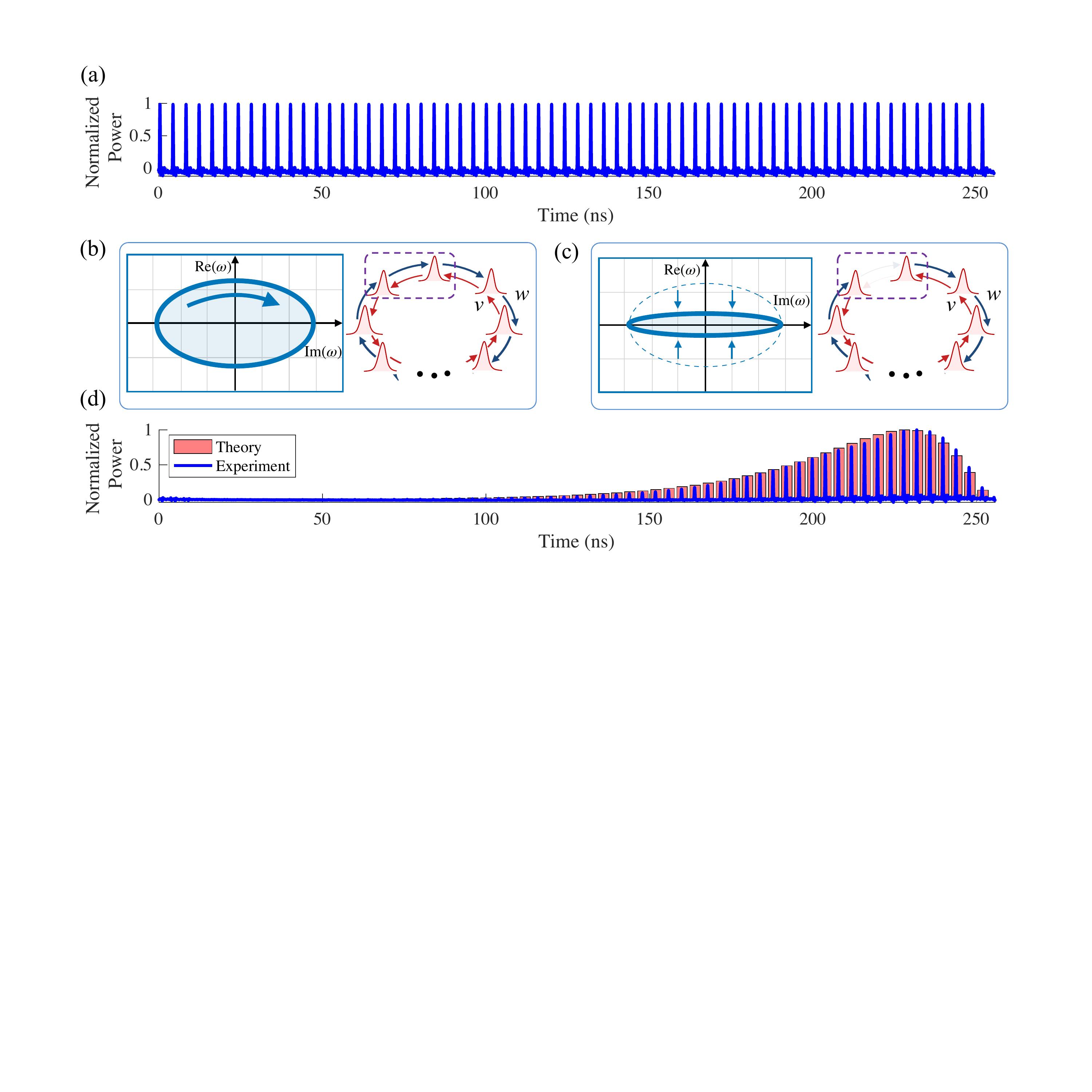}
    \caption{{\textbf{Non-Hermitian Topological Winding in a TTMLL}}~{\textbf{(a)}}~Measured mode-locked pulse pattern for an HN lattice with $w/v=\sqrt{2}$ and PBCs. Here the laser power is evenly distributed between all pulses in the cavity. \textbf{(b)}~Schematic illustration of the band diagram and pulse-to-pulse couplings of an HN lattice with PBCs. With PBCs, the energies $\omega$ of the HN model enclose a finite area in the complex energy plane and feature a nontrivial topological winding number. \textbf{(c)}~Schematic illustration of the band diagram and pulse-to-pulse couplings of an HN lattice with with imperfect OBCs. Note that, because the eigenvalues of the HN lattice change rapidly near the realization of exact OBCs~\cite{koch_bulk-boundary_2020}, the change in the eigenvalues is exaggerated for illustrative purposes. \textbf{(d)}~Measured mode-locked skin mode of an HN lattice with $w/v=\sqrt{2}$ and suppressed coupling between the first and final pulses. The theory is a best-fit skin mode that assumes a coupling ratio of $w/v=\sqrt{2}$ and treats the degree of coupling between the first and final sites as a fit parameter. Temporal mode-locking in this nascent skin mode provides evidence of the topological winding number in the lattice with PBCs.}
    \label{fig:hn_boundary_conditions}
\end{figure*}

We next study how changing the boundary conditions of an HN lattice influences the steady state of our topological MLL. In theory, varying the boundary conditions of our topological MLL from PBCs to perfect OBCs enables us to verify the correspondence between the bulk topological winding number and the NHSE that occurs in the presence of open boundaries~\cite{xiao_non-hermitian_2020}. However, in practice, the finite extinction ratio of our delay-line IMs limits our ability to perfectly suppress the coupling between the first and final sites of our synthetic HN lattice. Nonetheless, even with imperfect OBCs, we expect to see the emergence of a nascent skin mode, which presages the existence of the NHSE in the lattice with exact OBCs. Therefore, observing topological temporal mode-locking in this nascent skin mode still provides clear evidence of the nontrivial topological winding number in the lattice with PBCs.

We begin this experiment by preparing the couplings of our TTMLL to implement the HN model with PBCs and $w/v=\sqrt{2}$, and we tune our laser to operate with a small signal gain of $\sim\!22.2~{\rm dB}$ and an average output power of $\sim\!23~{\rm \mu{W}}$. In the presence of PBCs, we observe a uniform pulse train, as we expect from discrete translation symmetry [Figs.~\ref{fig:hn_boundary_conditions}\textcolor{red}{(a)}]. Next, we suddenly reduce the coupling at the boundary of our synthetic lattice. In this case, the laser undergoes a series of relaxation oscillations before the dissipative HN couplings and slow gain saturation temporally mode-locks the laser in the lowest-loss nascent skin mode of the underlying HN lattice. In Figs.~\ref{fig:hn_boundary_conditions}\textcolor{red}{(d)} we show an experimentally measured pulse pattern along with the best-fit lowest-loss state of an HN lattice with $w/v=\sqrt{2}$ and variable boundary conditions. We find that the best fit occurs when the coupling between the first and final lattice sites is suppressed by a factor of $\sim\!11.6$ [see Supplementary Information Sec.~4], which is reasonable given that our delay line IMs have a nominal DC power extinction ratio of $22~{\rm dB}$. Based on the formation of this nascent skin mode, we conclude that our topological MLL indeed hosts a nontrivial topological winding number $\mathcal{W}$ when the couplings are programmed to implement an HN lattice with PBCs.

\begin{figure*}
    \centering
    \includegraphics[width=\textwidth]{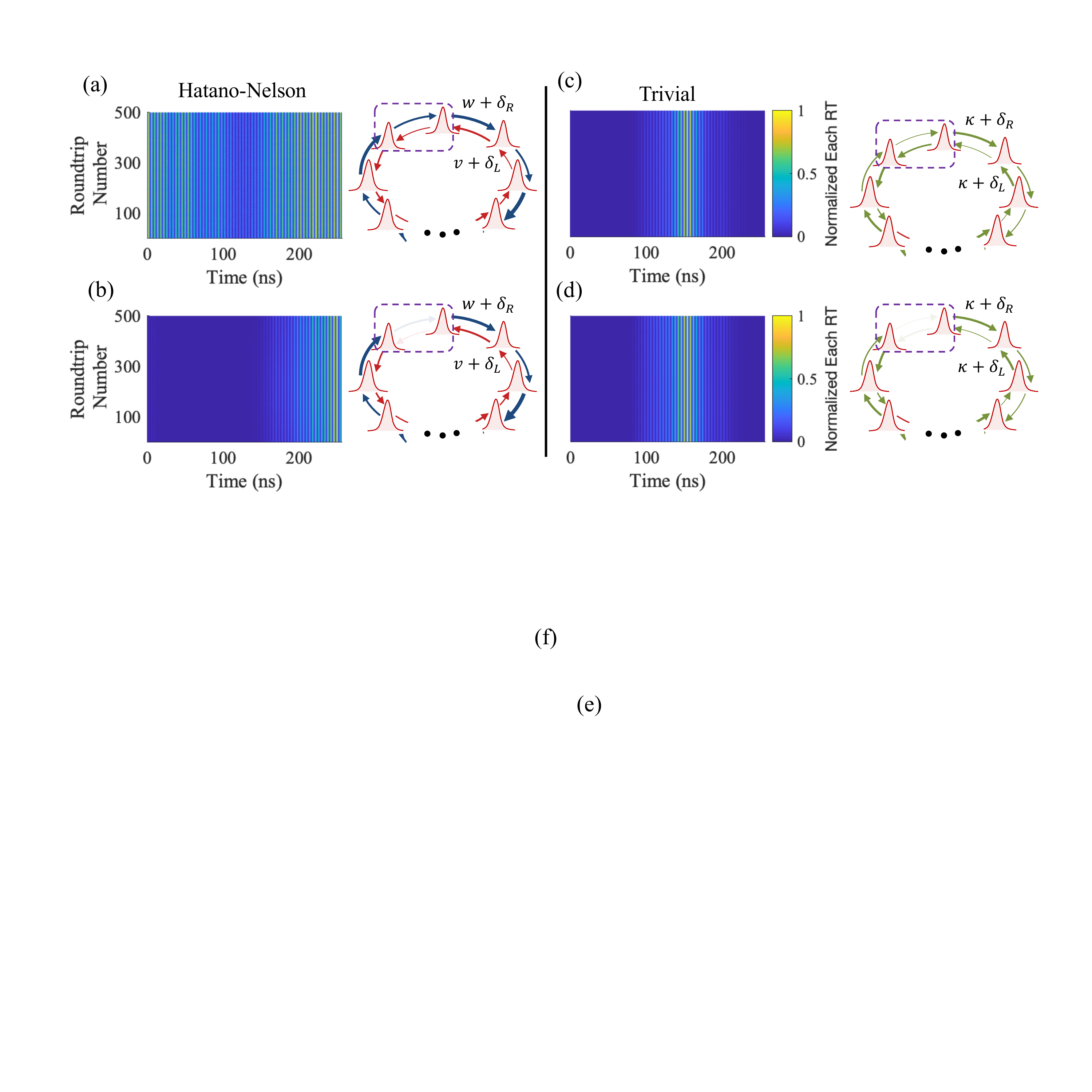}
    \caption{\textbf{Topological Robustness in the TTMLL} \textbf{(a)}~Mode-locked pulse pattern of an HN lattice with $w/v=4$, PBCs, and non-Hermitian coupling disorder distributed according to $\operatorname{Unif}(0,w/5)$. The topologically nontrivial winding number of the lattice confers robustness against disorder-induced localization. \textbf{(b)}~Mode-locked pulse pattern of the HN lattice from \textbf{(a)} with the coupling suppressed at the boundary. The observed signature of the NHSE indicates the existence of a nontrivial winding number in the lattice with PBCs. \textbf{(c)}~Localized mode-locked pulse pattern of a trivial lattice (coupling $\kappa$) with PBCs and the same level of non-Hermitian coupling disorder. \textbf{(d)}~Mode-locked pulse pattern of the corresponding trivial lattice with boundary coupling suppressed, which shows the absence of the NHSE in the trivial lattice. The pulses in the heat maps are artificially broadened to improve visibility.}
    \label{fig:disorder}
\end{figure*}

Given that our topological MLL can host a nontrivial topological winding number, we expect it to also exhibit topological robustness. In an HN lattice with PBCs, the existence of a nonzero topological winding number confers robustness against disorder-induced localization~\cite{gong_topological_2018}. In particular, when sufficiently small amounts of disorder are added to an HN lattice, either on-site or in the site-to-site couplings~\cite{hatano_vortex_1997}, the nontrivial winding number enables the states of the system to remain in a delocalized phase~\cite{gong_topological_2018}. This stands in sharp contrast to trivial 1D Hermitian lattices, where, in the thermodynamic limit, any finite level of disorder can produce Anderson localization~\cite{longhi_spectral_2021}. The ability to control the boundary conditions of the synthetic HN lattice in our topological mode-locked laser provides a distinct opportunity to study this robustness against disorder and to experimentally connect it to the existence of the topological winding number.

To experimentally demonstrate that the mode-locked temporal state of a disordered HN lattice with PBCs can remain in the delocalized phase in our topological MLL, we show that this state is highly sensitive to the boundary conditions on the HN lattice. This sensitivity has previously been established as a signature of the delocalized phase in point-gap topological systems~\cite{hatano_vortex_1997,hatano_non-hermitian_1998,xiong_why_2018}, and it can be correlated with other measures of delocalization~[see Supplementary Section 10]. We begin our experiments by programming the couplings of our topological temporal mode-locked laser to implement an HN lattice with $w/v=4$ and PBCs.  To each coupling, we add disorder distributed according to the uniform distribution $\operatorname{Unif}(0,w/5)$, so that we implement the HN model with 20\% coupling disorder. This off-diagonal disorder is distinctly non-Hermitian, and it stands in stark contrast to recent experimental realizations of non-Hermitian disorder, which have only considered on-site gain and loss~\cite{weidemann_coexistence_2021}. In the presence of 20\% coupling disorder and PBCs~[see Fig.~\ref{fig:disorder}\textcolor{red}{(a)}], the mode-locked pulse pattern of our TTMLL appears spread across the HN lattice. However, if we sharply reduce the coupling at the boundary, we observe that the power in the TTMLL concentrates near one boundary of the lattice. While the observed  nascent skin mode is again broadened due to residual coupling between the first and final pulses, the signature of the NHSE is unambiguous. The emergence of a nascent skin mode in Fig.~\ref{fig:disorder}\textcolor{red}{(b)} not only verifies that the mode-locked state remains in the delocalized phase in the presence of disorder but also establishes that a nontrivial winding number persists in the disordered HN lattice with PBCs~\cite{zhang_correspondence_2020}. Additional realizations of this phenomenon for other instances of disorder are presented in Supplementary Information Sec.~10.

We contrast our observations of disordered HN lattices with those of topologically trivial lattices with the same level of disorder. The mode-locked pulse patterns arising from one instance of disorder in the trivial lattice are shown in Fig.~\ref{fig:disorder}\textcolor{red}{(c,d)}. Unlike in our HN lattices, in the presence of PBCs [Fig.~\ref{fig:disorder}\textcolor{red}{(c)}], we observe a localized pulse pattern in the trivial lattice. Furthermore, when we reduce the coupling at the boundary [Fig.~\ref{fig:disorder}\textcolor{red}{(d)}], we observe that this mode-locked state remains largely unchanged, suggesting that it is not sensitive to the boundary conditions. This confirms the absence of a topologically nontrivial winding number in the trivial lattice and verifies that the trivial mode-locked state enters a localized phase.

While the contrast between Figs.~\ref{fig:disorder}\textcolor{red}{(a,b)} and Figs.~\ref{fig:disorder}\textcolor{red}{(c,d)} provides strong evidence that our HN lattice is topologically robust against localization, for other realizations of the disordered trivial lattice, the mode-locked pulse pattern can sometimes also appear sensitive to the boundary conditions  [see Supplementary Information Sec.~10]. For the trivial lattice, this apparent sensitivity to the boundary conditions is due to the relative weakness of the disorder and the small size of our synthetic lattice.  To remove any ambiguity caused by the finite size of our lattice, in Supplementary Information Sec.~10, we simulate the steady-state pulse patterns of the HN and trivial lattices as a function of the lattice size. Plotting a generalized inverse participation ratio for the two lattices as a function of the lattice size reveals that mode-locked state of the HN lattice remains in a delocalized phase, while that of the trivial lattice becomes localized. 

While the present experiments focus on temporal mode-locking in topological systems, the architecture of the TTMLL may be straightforwardly adapted to study other fundamental phenomena as well as potential applications. The time-multiplexed architecture of our dissipative couplings enables the realization of multiple synthetic dimensions and arbitrary pulse-to-pulse couplings~\cite{leefmans_topological_2022}. This capability may enable our TTMLL to implement higher-dimensional NHSEs~\cite{song_two-dimensional_2020}, point-gap topological models with long-range couplings~\cite{wang_generating_2021}, non-Hermitian Anderson topological insulators~\cite{lin_observation_2022}, and other coupled systems of fundamental interest, such as those that exhibit $\mathcal{PT}$-symmetry~\cite{ozdemir_paritytime_2019}. Beyond fundamental science, one can imagine applying a temporally mode-locked laser (TML) to sensing and optical computing~\textemdash~for example, by designing the intracavity couplings so that the mode-locked temporal structure is highly sensitive to changes in the couplings, by using the steady-state of the TML in the low-power regime to solve unwieldy eigenvector problems, or by incorporating the dynamics of the TML into a time-multiplexed computing architecture such as the coherent optical Ising machine~\cite{marandi_network_2014}.

Additionally, while the topological temporal mode-locking mechanism studied in this work utilizes active mode-locking and a slow, solid-state laser gain, the architecture of the TTMLL may be adapted to study topological phenomena in other mode-locked photonic systems~\textemdash~including various kinds of passively mode-locked lasers~\cite{haus_additive-pulse_1994,keller_recent_2003}; fast-gain, semiconductor mode-locked lasers~\cite{rafailov_mode-locked_2007}; cavity solitons~\cite{kippenberg_dissipative_2018,lu_synthesized_2021,englebert_temporal_2021}; and sychronously pumped optical parametric oscillators~\cite{marandi_network_2014,roy_topological_2022}. Indeed, the notion of combining a nonlinear pulse formation mechanism with pulse-to-pulse intracavity couplings may be adapted to any source that generates a uniform, intracavity pulse train, as such a pulse train may be thought as a temporal synthetic lattice whose pulses can be coupled with optical delay lines. In this sense, our realization of topological temporal mode-locking reveals myriad opportunities to study the interplay between topology and a diversity of mode-locked resonators and has the potential to expand the scope of nonlinear and non-Hermitian topological photonics. 

In conclusion, we have theoretically proposed and experimentally demonstrated the topological temporally mode-locked laser. By harnessing the interplay between a slow gain saturation nonlinearity and intracavity dissipative couplings, we observed a nonlinearity-driven NHSE and, by adding non-Hermitian coupling disorder to an HN lattice with PBCs, we demonstrated the topological robustness of the HN model against disorder-induced localization. We explained how our topological MLL can be extended to study both different fundamental phenomena and potential applications, and we discussed how our experimental architecture may be adapted to study nonlinear and non-Hermitian topological phenomena with other short-pulse sources. In future work, we will extend our experiments to both explore topological temporal mode-locking in the high-power regime, where the Kerr nonlinearity plays a more substantial role, and to study topological dynamics in other mode-locked photonic resonators.

\section*{Acknowledgments}
The authors are grateful to Kerry Vahala and Lue Wu for lending equipment useful to this work. The authors acknowledge support from  NSF Grants No. 1846273 and 1918549 and AFOSR Award No. FA9550-20-1-0040. F.N. acknowledges support from ARO (W911NF-18-1-0358), JSPS (JP20H00134), AOARD (FA2386-20-1-4069), and FQXi (FQXi-IAF19-06). The authors wish to thank NTT Research for their financial and technical support.

\bibliography{bibliography.bib}
\bibliographystyle{mynaturemag}

\section*{Methods}

\subsection*{Experimental Setup and Calibration}

As discussed in the main text, the experimental platform utilized in this work is a topological temporally  mode-locked laser (TTMLL) [Fig.~\ref{fig:schematic}\textcolor{red}{(a)}]. This laser architecture consists of a main laser cavity and two optical delay lines, which introduce nearest-neighbor couplings between the pulses in the laser. For a more detailed picture of our setup, please see Fig.~\textcolor{red}{1} in Supplementary Information Sec.~1.

Our TTMLL is built from polarization-maintaining fiber patch cables and discrete optical components. All components and patch cables are terminated with FC/APC connectors to reduce back-reflections, and they are joined with PM fiber mating sleeves. The primary optical elements of the main cavity are an erbium-doped fiber amplifier (EDFA) and an intensity modulator (IM), which together are responsible for pulse formation in the laser cavity. The main optical element in each delay line is an IM, which controls the coupling strengths between the pulses. In addition to these fiber elements, there are free space delays in both the main cavity and in the two optical delay lines, which provide the flexibility necessary to properly match the lengths of the different lines.

We set the lengths of the main cavity and the delay lines by injecting the pulses of an auxiliary mode-locked laser into our system. We tune the length of the main cavity so that these auxiliary pulses are resonant within the cavity, and we tune the delay line lengths so that they couple the auxiliary pulses.

Next, we use the auxiliary pulses to perform a preliminary calibration of the delay line IM responses. We disconnect the feedback of the main cavity and reconfigure the path to our photodetector so that we observe only the throughput of one optical delay line. We then linearly sweep the RF voltage applied to IM in the delay line and observe the resultant power of the auxiliary pulses. With this data, we construct an applied voltage versus optical power curve, which we use to generate the coupling waveforms for our experiments. 

After this calibration step, we reconnect the feedback in the main cavity.

Because the lengths of our TTMLL's paths and the responses of the delay line modulators are calibrated with an auxiliary MLL, we want the pulses generated in our TTMLL to have the same repetition rate as the auxiliary pulse train and to temporally overlap with the positions of the auxiliary pulses in the cavity. The first condition ensures that the repetition rate of the pulses is compatible with the cavity length, while the second ensures that the pulses of our TTMLL see the same response function of the delay line IMs.

To match the repetition rate of our TTMLL with that of the auxiliary MLL, we use a 10~MHz reference from our auxiliary MLL as a clock for the RF function generator responsible for actively mode-locking the main cavity. We then set the frequency of the RF function generator's sinusoidal output to be the known repetition rate of the auxiliary laser.

To match the temporal overlap of the TTMLL's pulses with those of the auxiliary laser, we place an RF phase shifter at the output of the RF function generator. With the RF output on and the auxiliary pulses inside the cavity, we tune the phase of the RF output until we maximize the resonant power in the laser cavity. At this point we reason that the TTMLL's pulses will be generated in the same temporal positions as the auxiliary pulses.

We now remove the auxiliary pulses from the laser cavity.

All of the calibration steps described above occur below threshold. In an effort to ensure that our calibration does not break down once we initiate lasing in our TTMLL, we perform a second round of calibration procedures above threshold.

First, we disconnect the fiber that recombines the delay lines with the main cavity, so that pulses can pass through the delay lines without coupling back into the main cavity. We reconfigure the path to our detector so that we observe the pulses passing through one of the delay lines.

We now bring the laser cavity above threshold, and we tune the length of the main cavity slightly until we observe a stable pulse train. Then, using our initial calibration for the delay line IMs, we generate the coupling waveforms that we want to use for a particular experiment, and we apply them to the delay line IMs. We view the throughput of the delay lines on our detector and evaluate whether it agrees with the expected couplings. If the throughput does not agree with our expectations, we further tune the positions of the optical pulses until the agreement with the expected throughput improves. If small discrepancies persist after this tuning, we manually update the coupling waveform to achieve the desired delay line throughput.

Note that the calibration procedure described here is only valid for a limited amount of time, as the length of our main cavity is not locked to the RF function generator that produces the optical pulses. Therefore, the relative drift of the main cavity and the RF function generator can both destabilize the pulse train in the main cavity and change the pulse positions, which can modify the intracavity coupling in our experiment in unexpected ways. In the future, regenerative mode-locking~\cite{le_nguyen_binh_ultra-fast_2011} may help to improve the stability and consistency of our system.

\subsection*{Experimental Procedure}

To initiate our experiments, we first block the light in the two delay lines by placing beam blocks in their free space delays. We bring the main cavity above threshold and tune the length of the main cavity and the gain of the EDFA until we observe a train of uniformly spaced pulses. Then we unblock the delay lines one at a time and attempt to lock them to constructively interfere with the main cavity. As the threshold of our TTMLL is lower with the delay lines constructively locked to the main cavity, we next reduce the gain in our laser to lower the intracavity power to the point where we can neglect the effects of the Kerr nonlinearity.

In this initial operating phase, the delay lines implement constant couplings between all of the pulses. However, while these couplings are constant, they need not be the same for each delay line. In particular, by tuning the coupling strength in one delay line to be greater than that in the other, we can immediately initialize our laser in the topological phase of the HN model. Then we can study the non-Bloch bulk-boundary correspondence in the HN model by reducing the coupling between two pulses to achieve a boundary.

Next, we change the coupling waveforms applied to the delay line IMs to achieve topological temporal mode-locking. We implement these couplings for several minutes at a time, and we observe the pulse pattern produced by the couplings on an oscilloscope. To ensure that the observed pulse pattern occurs in the expected position in the synthetic lattice of the laser pulses, we reuse the auxiliary MLL from our calibration. By sending one pulse from this laser directly to a second detector on each roundtrip of our TTMLL, we produce a reference point from which to determine the positions of the pulses in our TTMLL relative to the applied couplings. This positioning procedure is detailed in Supplementary Information Sec.~5.

Finally, we manually stop the oscilloscope and save the data on-screen to a USB drive.

\pagebreak

\begin{figure*}
    \centering
    \includegraphics[width=\textwidth]{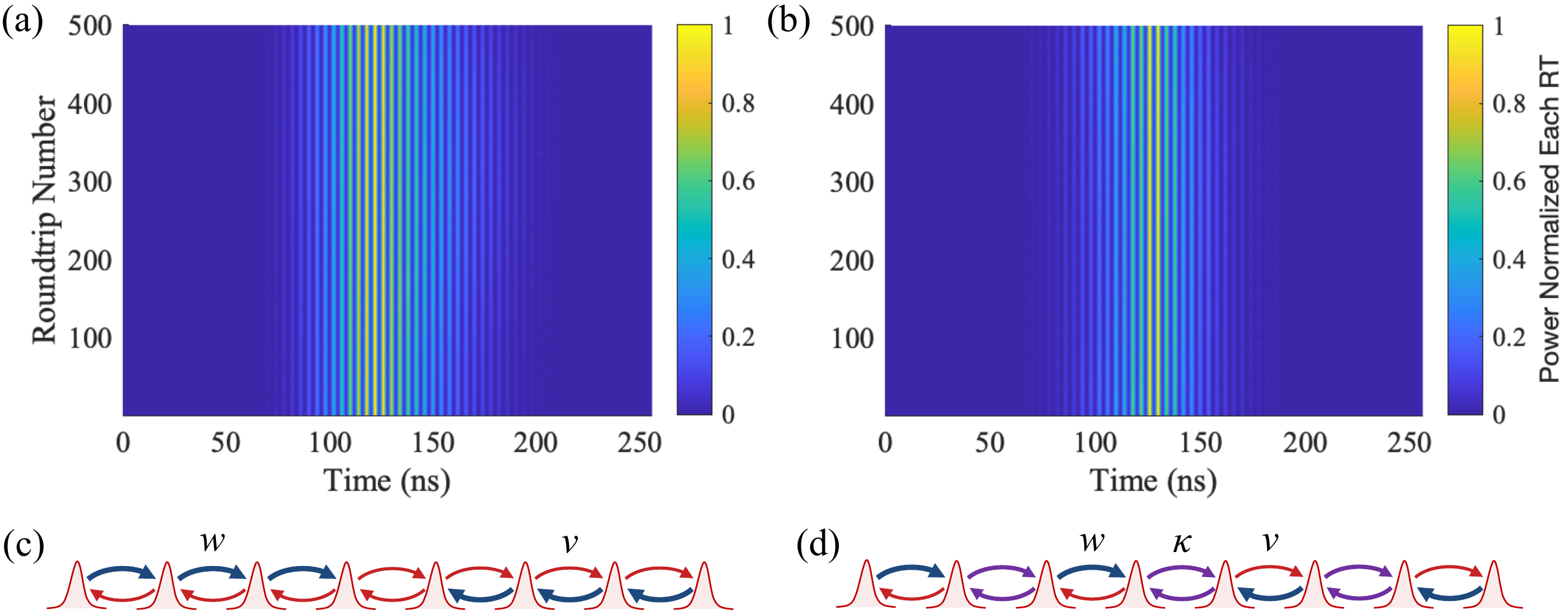}
    \caption*{Extended Data Figure 1: \textbf{Additional Data for Topological Mode-Locking at HN and NH-SSH Domain Walls} \textbf{(a)}~Temporal mode-locking in the HN domain-wall skin mode for 500 roundtrips. \textbf{(b)}~Temporal mode-locking in the NH-SSH domain-wall skin mode for 500 roundtrips. The plots shown in Fig.~\ref{fig:domain_walls}\textcolor{red}{(a,d)} are produced using \textbf{(a)} and \textbf{(b)}, respectively, together with the averaging procedure discussed in Supplementary Information Sec.~4. \textbf{(c)}~Schematic representation of the HN domain wall. \textbf{(d)}~Schematic representation of the NH-SSH domain wall. Note that the pulses in these figures are artificially broadened for visibility.}
    \label{fig:extended_domain_walls}
\end{figure*}

\begin{figure*}
    \centering
    \includegraphics[width=0.8\textwidth]{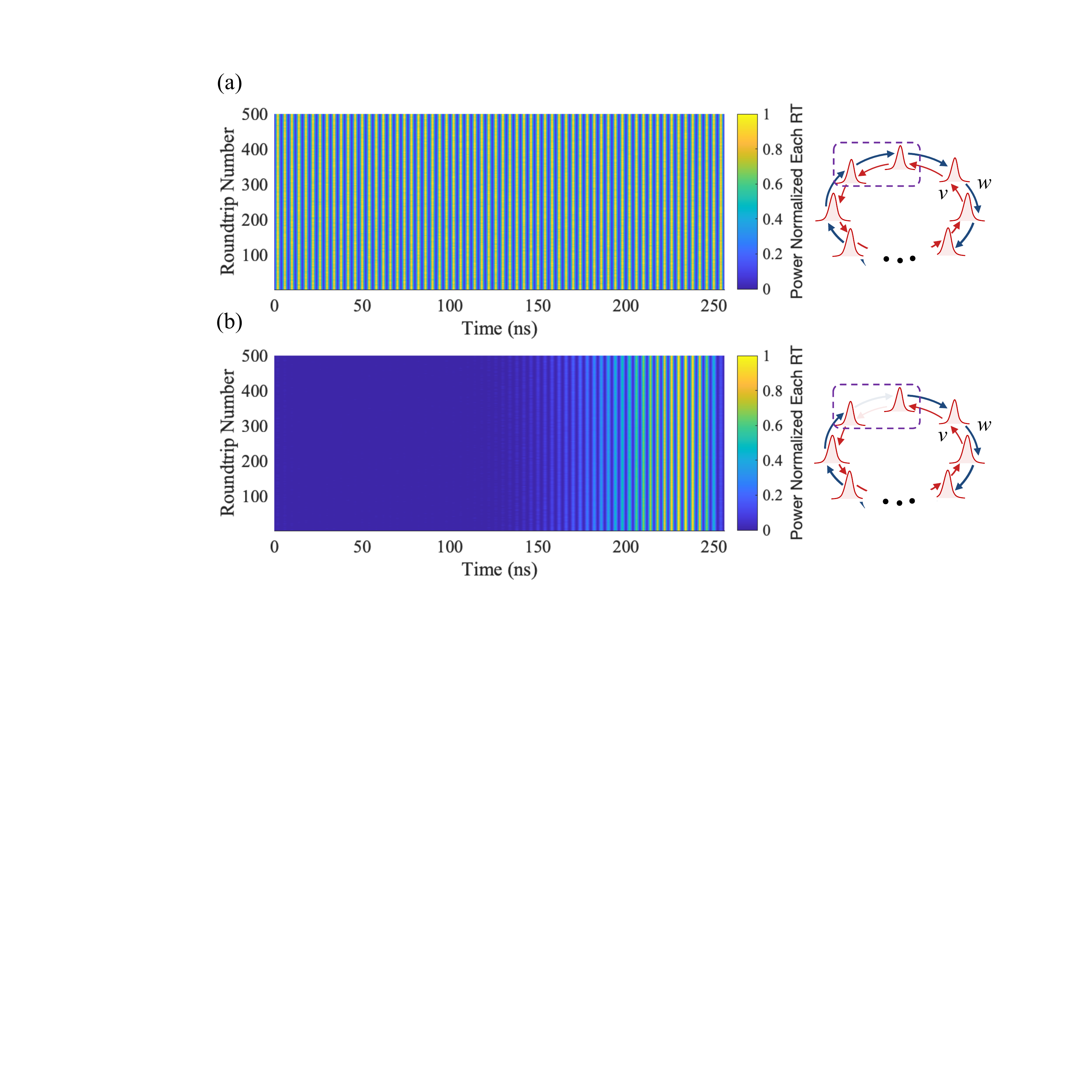}
    \caption*{Extended Data Figure 2: \textbf{Additional Data for Non-Hermitian Topological Winding in a TTMLL} \textbf{(a)}~Topological temporal mode-locking of the HN lattice with PBCs for 500 roundtrips. \textbf{(b)}~Topological temporal mode-locking in the lowest-loss mode of the HN lattice with reduced coupling between the first and final lattice sites. The traces shown in Fig.~\ref{fig:hn_boundary_conditions}\textcolor{red}{(a,d)} are generated from \textbf{(a)} and \textbf{(b)}, respectively, using the averaging procedure described in Supplementary Information Sec.~4. Once again, the pulses in these figures are artificially broadened for visibility.}
    \label{fig:extended_phase_transition}
\end{figure*}

\pagebreak

\end{document}